\documentclass[aps,prl,preprint,superscriptaddress]{revtex4-1}

\usepackage{graphicx}
\usepackage{color}
\usepackage{SIunits}

\begin{document}

\title{Strong Group Velocity Dispersion Compensation\\
       With Phase-Engineered Sheet Metamaterials}

\author{Babak Dastmalchi}
\email{babakds@ameslab.gov}
\affiliation{Ames Laboratory---U.S. DOE and Department of Physics and Astronomy, Iowa State University, Ames, Iowa 50011, USA}
\affiliation{Institute of Electronic Structure and Lasers (IESL), FORTH, 71110 Heraklion, Crete, Greece}
\affiliation{Center for Surface and Nanoanalytics, Johannes Kepler University, Altenberger Str. 69, 4040 Linz, Austria}
\author{Philippe Tassin}
\affiliation{Ames Laboratory---U.S. DOE and Department of Physics and Astronomy, Iowa State University, Ames, Iowa 50011, USA}
\affiliation{Department of Applied Physics, Chalmers University, SE-412 96 G\"{o}tenborg, Sweden}
\author{Thomas Koschny}
\affiliation{Ames Laboratory---U.S. DOE and Department of Physics and Astronomy, Iowa State University, Ames, Iowa 50011, USA}
\author{Costas M. Soukoulis}
\affiliation{Ames Laboratory---U.S. DOE and Department of Physics and Astronomy, Iowa State University, Ames, Iowa 50011, USA}
\affiliation{Institute of Electronic Structure and Lasers (IESL), FORTH, 71110 Heraklion, Crete, Greece}

\begin{abstract}
Resonant metamaterials usually exhibit substantial dispersion, which is considered a shortcoming for many applications. Here we take advantage of the ability to tailor the dispersive response of a metamaterial introducing a new method of group-velocity dispersion compensation in telecommunication systems. The method consists of stacking a number of highly dispersive sheet metamaterials and is capable of compensating the dispersion of optical fibers with either negative or positive group-velocity dispersion coefficients. We demonstrate that the phase-engineered metamaterial can provide strong group-velocity dispersion management without being adversely affected by large transmission loss, while at the same time offering high customizability and small footprint.
\end{abstract}

\maketitle

Dispersion management is an indispensable element of optical communication systems, where dispersive effects---originating from the materials, waveguide (fiber) geometries, and optical amplification---accumulate to set limits on both the distance and the bit rate of the data transfer. Various compensation schemes have been developed to manage group-velocity dispersive effects~\cite{Agrawal2010fiber,ramachandran2007fiber}, but fundamental limits on integrability, footprint and customizability are imposed by the physics in contemporary dispersion management systems. Recent advances in nanofabrication and breakthroughs in the field of metamaterials~\cite{Smith2004,Engheta2006,Shalaev2007,Liu2011,Soukoulis2011,Zheludev2012} have opened up a new range of possibilities in device development. Most metamaterials rely on highly resonant structures that force light to undergo a large phase change near resonance frequencies. This results in strong dispersion in a narrow spectral range, making them suitable for dispersion management purposes. Indeed, it was recently shown that light passing through a so-called metasurface experiences up to a $2\pi$ phase shift upon transmission/reflection in a system that is much thinner than the free-space wavelength of the incident light, mimicking a phase discontinuity~\cite{Yu2011,Aieta2012}. Phase properties in such metasurfaces are shown to be easily tailorable, although it is important to note that this comes at the cost of absorption in the dispersive region.

In this Letter, we show that dispersion-engineered metamaterials exhibiting a classical analogue of electromagnetically induced transparency (EIT) can address the group-velocity dispersion problem without being adversely affected by return loss. Originally, EIT is a quantum mechanical phenomena characterized by a narrow transmission window in a relatively wide absorption band~\cite{Harris1990,Harris1997,Fleischhauer2000,Matsko2001,Mandel2001,Fleischhauer2005}. The change in the transmission is accompanied by a strongly nonlinear dispersion relation and, hence, by group velocity dispersion (GVD). Several groups have now demonstrated that the phenomena can be reproduced in purely classical systems by way of metamaterials \cite{Fedotov2007,Zhang2008,Tassin2009,Yannopapas2009,Liu2009,Han2011,Jin2011,Verslegers2012}. In classical EIT, the transmission amplitude, the bandwidth and the center frequency of the transparency window can be modified through the geometry and the constituent material properties.

\begin{figure}[t!]
\includegraphics[width=8cm]{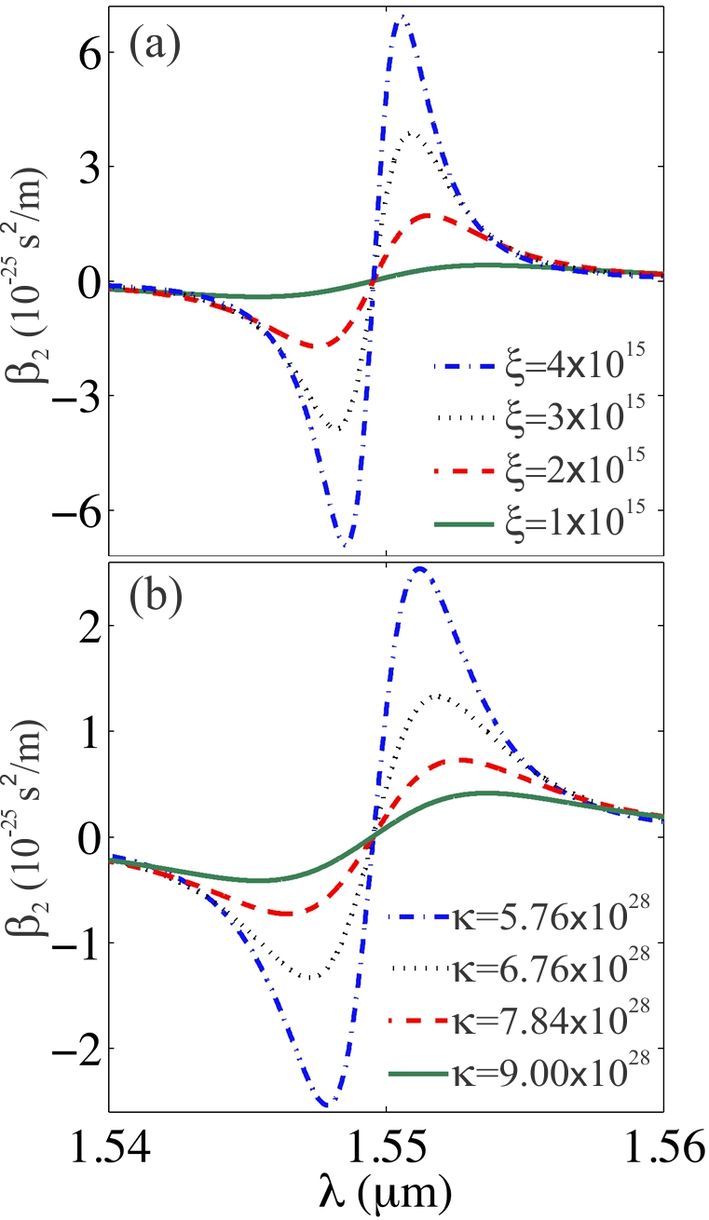}
\caption{(a) Variation of GVD with regard to the changes in the static susceptibility $\xi=\epsilon_0\chi_{se}^{static}$ with $\kappa=9.00\times10^{28}$.  (a) Variation of GVD with regard to the changes in the coupling efficiency $\kappa$ with $\xi=1\times10^{15}$. $\gamma_\mathrm{r}=0.01\times10^{15}$ and $\gamma_\mathrm{d}=0.000001\times10^{15}$ in both figures.}
\label{fig:param_scan}
\end{figure}

\begin{figure}[t!]
\includegraphics[width=12cm]{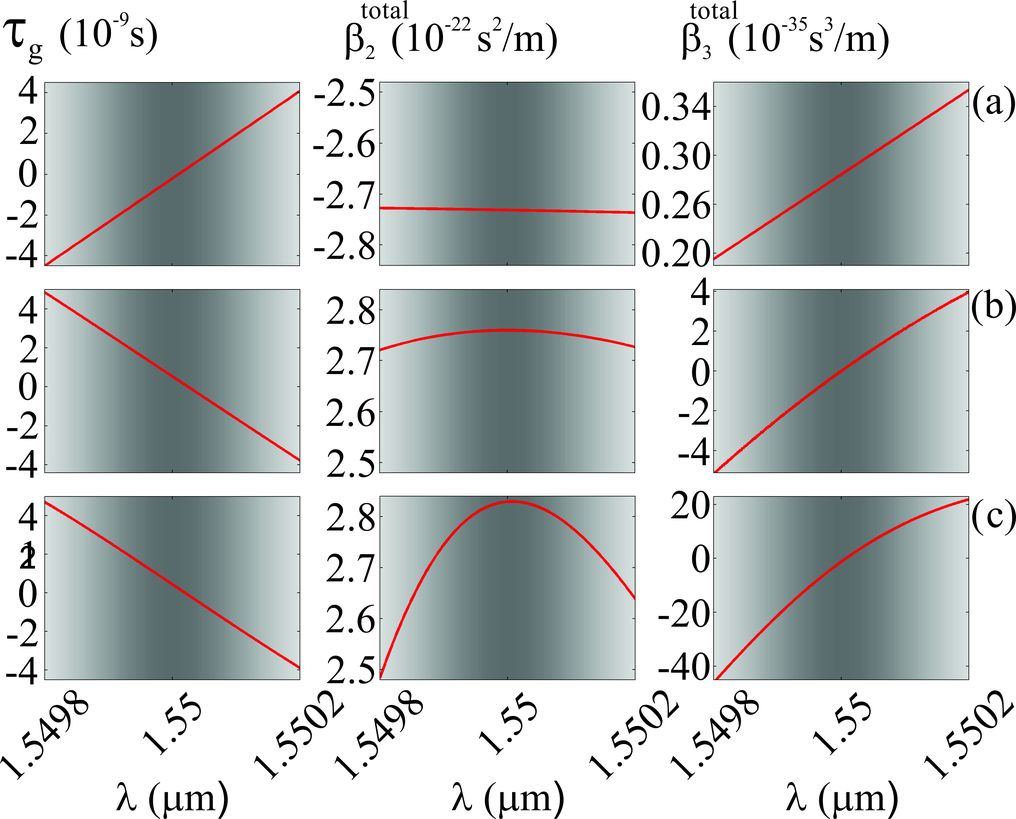}
\caption{(a) The change in the group delay, second and third order total residual dispersion for (a) 25 km long fiber (b) 908 EIT layers with $\xi=1.5\times10^{15}$ and (c) 130 EIT layers with $\xi=4\times10^{15}$. Gray shaded area shows the spectral distribution of the Gaussian pulse with $\tau_0=10~ps$. Resonance frequency of both EIT models is chosen to reach to the maximum dispersion at the center frequency of the Gaussian pulse.}
\label{fig:disp}
\end{figure}

Let us briefly state the problem in a quantitative way. In telecommunication systems, data is transmitted as a sequences of pulses of certain shape and width, formed by superposition of frequency-dependent plain waves with a particular weight function, e.g., Gaussian. For a narrow-banded pulse, the propagation constant $\beta$ can be expanded around the center frequency $\omega_0$:
\begin{equation}
\label{eq:expansion}
\beta = \beta_0+\beta_1 \Delta\omega+\beta_2\Delta\omega^2+\beta_3\Delta\omega^3+\ldots,\\
\end{equation}   
where 
\begin{equation}
\label{eq:expansion1}
\beta_i=\frac{\partial^i \beta}{\partial\omega^i}|_{\omega=\omega_0},\quad i=1,2,3,\ldots\\
\end{equation}    
$\beta_1=\frac{1}{v_\mathrm{g}}$ is the invserse of the group velocity. $\beta_2$, known as the group-velocity dispersion (GVD) coefficient, manifests itself as a broadening in the pulse width. The higher-order terms result in distortion of the pulse from the initial form, but are usually negligible due to small strength. Pulse broadening, however, causes inter-symbol interference and limits the bit rate of the communication line. To achieve large distance communication, it is necessary to restore the data pulses to the original width using a dispersion compensation method. For this purpose, the broadened pulse is usually sent through a second medium with the opposite sign of dispersion. Indeed, the effect of GVD can be cancelled completely if the lengths and magnitudes of the second-order dispersion coefficients for the two media satisfy the following condition:
\begin{equation}
\label{eq:condition}
L_1\beta_2^\mathrm{M1}+L_2\beta_2^\mathrm{M2}=0,
\end{equation}    
where $L_i$ and $\beta_2^{\mathrm{M}i}$ are the length and GVD of the $i^\mathrm{th}$ medium, respectively~\cite{Agrawal2010fiber}. $L_i\beta_2^\mathrm{Mi}$ quantifies the total residual dispersion imposed on the pulse after travelling through the $i^\mathrm{th}$ medium.
\begin{figure}[t!]
\includegraphics[width=8cm]{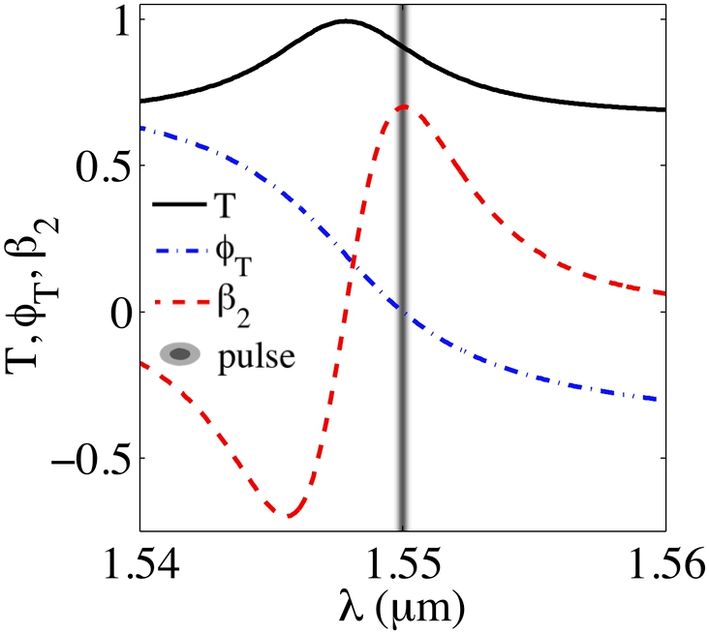}
\caption{Transmission amplitude (black), transmission phase (blue), and group-velocity dispersion (red) for a generic EIT system. The gray line shows the spectral distribution of a Gaussian pulse of the form $\hat{E}(\omega-\omega_0) = \sqrt{2\pi\tau_0^2}e^{-\frac{1}{2}\tau_0^2(\omega-\omega_0)^2}$ to be dispersion compensated. $\tau_0=\unit{10}{\pico\second}$, which corresponds to a bandwidth of \unit{0.18}{\nano\meter}, or a transmission rate of at least 10 Gbit/\second.}
\label{fig:generic}
\end{figure}
Fig.~\ref{fig:generic} shows the behavior of an EIT medium. The transmission phase and, hence, the GVD  changes drastically around the transmission peak. Interestingly, GVD has opposite signs around the transmission peak, which can be utilized to manage both negative and positive dispersion by positioning the pulse at the right- or left-hand side of the transmission peak. Theoretically, the mechanism underlying EIT can be best explained by a model of two interacting resonant modes, having nearly the same resonance frequency and differing in coupling strength to the incident electromagnetic field. One mode, known as the radiative or bright mode, can strongly couple to the external incident field, while the other, the dark mode, barely couples to the external field. The dark and bright modes can, nevertheless, interact with each other through near-field coupling, resulting in strongly resonant and dispersive behavior. It has been experimentally verified that EIT can be implemented as a thin conductive sheet of magnetic and electric meta-atoms~\cite{Tassin2012a}. Reflection and transmission of the conductive sheet can be connected to the electric and magnetic conductivity~\cite{Tassin2012}:
\begin{eqnarray}
\label{eq:retrievalinv}
\begin{array}{l}
		r=\frac{2(\zeta\sigma^\mathrm{(e)}_{||}-\zeta^{-1}\sigma^\mathrm{(m)}_{||})}{4+2\zeta\sigma^\mathrm{(e)}_{||}+2\zeta^{-1}\sigma^\mathrm{(m)}_{||}
			  	+\sigma^\mathrm{(e)}_{||}\sigma^\mathrm{(m)}_{||}},\\
		t=\frac{4-\sigma^\mathrm{(e)}_{||}\sigma^\mathrm{(m)}_{||}}{4+2\zeta\sigma^\mathrm{(e)}_{||}+2\zeta^{-1}\sigma^\mathrm{(m)}_{||}
			  	+\sigma^\mathrm{(e)}_{||}\sigma^\mathrm{(m)}_{||}}.	
\end{array}
\end{eqnarray}
In Eq.~(\ref{eq:retrievalinv}), $\sigma^\mathrm{(e)}_{||}$ ($\sigma^\mathrm{(m)}_{||}$) is the electric (magnetic) conductivity of the EIT sheet. To satisfy the compensation condition [Eq.~(\ref{eq:condition})], one can either adjust the dispersion of an individual sheet or use multiple EIT sheets. As it will be discussed below, multiple sheets might be necessary to perform an ideal dispersion compensation. In addition, metamaterials provide us the unique opportunity of impedance matching to avoid complications arising from multiple reflections~\cite{Pfeiffer2013}. Applying the impedance matching condition, $\zeta\sigma^\mathrm{(e)}_{||}=\zeta^{-1}\sigma^\mathrm{(m)}_{||}$, Eq.~(\ref{eq:retrievalinv}) simplifies to
\begin{eqnarray}
\label{eq:retrievalnoreflection}
\begin{array}{l}
		r=0,\\
		t=\frac{2-\sigma_{||}^\mathrm{(e)}}{2+\sigma_{||}^\mathrm{(e)}}.
\end{array}
\end{eqnarray}
The electric conductivity of a single sheet can be derived by solving a coupled-resonator model \cite{Tassin2012a}:
\begin{equation} 
	\sigma_{||}^\mathrm{(e)}=-\mathrm{i}\omega\xi\frac{D_\mathrm{d}(\omega)}{D_\mathrm{b}(\omega) D_\mathrm{d}(\omega)-\kappa^2},
\end{equation}
where $D_\mathrm{d}=\omega_\mathrm{d}^2-\mathrm{i}\gamma_\mathrm{d}\omega-\omega^2$ and $D_\mathrm{b}=\omega_\mathrm{b}^2-\mathrm{i}\gamma_\mathrm{b}\omega-\omega^2$.
$\gamma_\mathrm{d}$ and $\omega_\mathrm{d}$ ($\gamma_\mathrm{b}$ and $\omega_\mathrm{b}$) are, respectively, the damping factor and resonance frequency of the dark (bright) mode, and $\kappa$ denotes the near-field coupling strength of the two resonators. $\xi=\epsilon_0\chi_\mathrm{s}^\mathrm{static}$ is the static susceptibility of the conductive sheet. For a single EIT sheet, $\beta_2$ can be tailored either by adjusting the coupling strength or by changing $\xi$ (see Fig.~\ref{fig:param_scan}). The latter is proportional to the area density of packed resonators. It can be observed in Fig.~\ref{fig:param_scan} that increasing either of the parameters $\kappa$ or $\xi$ increases $\beta_2$ in favor of stronger dispersion compensation, although at the expense of reduced bandwidth of the dispersion curve.
The region close to the GVD extrema, defining the $\beta_2$ bandwidth, provides the highest dispersion compensation amplitude. Moreover, the third-order dispersion ($\beta_3$) has its lowest value in this region, although it increases when moving away from the extremal points. Therefore, to take advantage of strong dispersion compensation as well as low pulse distortion due to third-order dispersion, the data pulse should be accommodated well inside the flat band of $\beta_2$. This imposes a limit on the maximally achievable $\beta_2$ with a single EIT sheet since the bandwidth and magnitude of $\beta_2$, as shown in Fig.~\ref{fig:param_scan}, are complementary and cannot be maximized simultaneously. To satisfy Eq.~(\ref{eq:condition}) and avoid large pulse distortion at the same time, it is necessary to use multiple EIT sheets. However, it should be kept in mind that accumulated losses from multiple EIT sheets would sacrifice the amplitude of transmitted pulses in exchange for lower distortion. To further understand this trade-off, we have investigated two different configurations. The first case [Fig.~\ref{fig:disp}(b)] is designed to show a relatively small GVD in the spectral width of the Gaussian input pulse. The parameters for the individual sheets in this arrangement are $\xi=1.5\times10^{15}$, $\kappa=6.76\times10^{28}$, $\gamma_\mathrm{r}=0.01\times10^{15}$ and $\gamma_\mathrm{d}=0.000001\times10^{15}$. This configuration requires using 908 EIT units to satisfy the compensation condition~(\ref{eq:condition}). For the second case [see Fig.~\ref{fig:disp}(c)], $\beta_2$ for a single sheet is boosted by increasing $\xi$ to $4\times10^{15}$, resulting in a much more dispersive system (larger $\beta_2^\mathrm{total}$), but also in increased third-order dispersion in the spectral region of interest. This reduces the number of required sheets to 130. Figure~\ref{fig:disp} compares the group delay and the two lowest-order total residual dispersion of both cases when used to compensate the GVD of a single-mode step-index fiber. The fiber is considered to have a cylindrical core of radius $a$ and refractive index $n_\mathrm{c}$, and a cladding with a refractive index of $n$. The effective index of the fiber, considering both material and waveguide dispersion, is approximated by~\cite{Chang1979}:
\begin{equation} 
n_\mathrm{eff}=n\left(1+b\Delta\right)
\end{equation}
The approximation is valid for small index contrast ($\Delta=\frac{n_\mathrm{c}-n}{n}<<1$), where 
\begin{equation} 
\begin{array}{l}
    v=a k_{0} \sqrt{n_{\mathrm{c}0}^2-n^2}\\
    W=1.1428v-0.996\\
    b=\frac{W^2}{v^2}
\end{array}
\end{equation}
The fiber core diameter is chosen to be \unit{5.3}{\micro\meter}, $\Delta=0.006$ and $n$ is calculated from a Sellmeier model for quenched silica~\cite{Fleming1978}.
For the numerical calculations of the wave propagation, a transfer matrix method has been used. Individual EIT sheets are assumed to be decoupled and arranged in a periodic array embedded in the fiber medium. Reflection and transmission coefficient of EIT sheets are calculated from the model in Eq.~(\ref{eq:retrievalnoreflection}). It should also be noted that the dark and bright resonance frequencies are adjusted so that the GVD peak coincides with the center frequency of a Gaussian data pulse of the form $\hat{E}(\omega-\omega_0) = \sqrt{2\pi\tau_0^2}e^{-\frac{1}{2}\tau_0^2\omega^2}$, with $\tau_0=\unit{10}{\pico\second}$. 
The dispersion compensated pulse, using the first configuration [Fig.~\ref{fig:disp}(a)], has an electric field amplitude of $|E|=0.0022$, whereas, for the second configuration [Fig.~\ref{fig:disp}(c)], the amplitude is $|E|=0.639$, showing almost three orders of magnitude improvement with respect to the first case. The substantial reduction in transmission loss should be attributed only to the avoidance of dissipative losses in the individual EIT sheets, since the radiative loss is eliminated by using impedance matching. Theoretically, loss can be further reduced by decreasing the dark and bright resonator damping. For purposes of illustration, we have calculated this limit of vanishing damping losses ($\gamma_\mathrm{d}=0$ and $\gamma_\mathrm{r}=0$) and we have found that the pulse amplitude as well as the pulse width can be restored near to its initial value given the higher-order dispersions can be minimized arbitrarily by designing a flatter $\beta_2$ extremum and using more EIT sheets.

\begin{figure}[Ht]
\includegraphics[width=12cm]{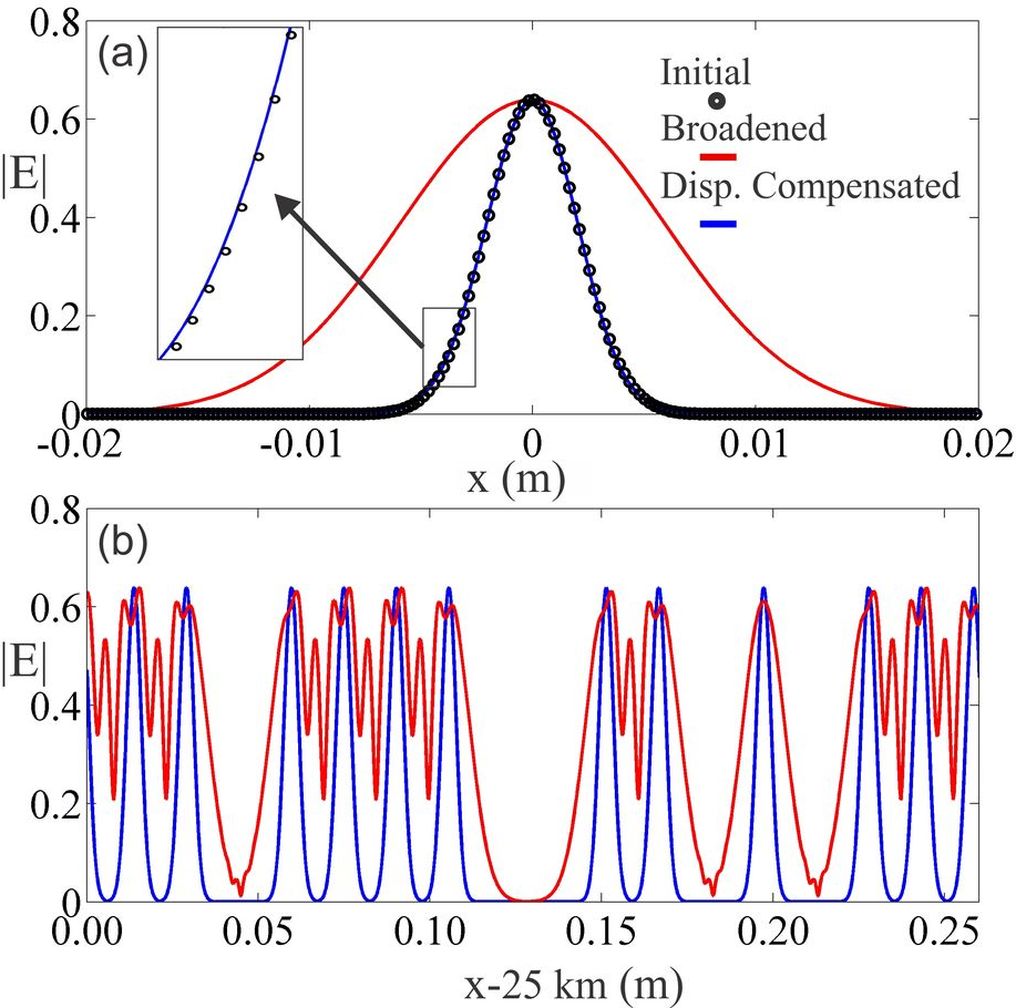}
\caption{(a)~Comparison of broadened, compressed and initial pulses for 130 EIT sheets with $\xi=4\times10^{15}$. Inset shows the deviation from the initial Gaussian shape due to higher order dispersion. (b)~A train of broadened Gaussian pulses at the end of a \unit{25}{\kilo\meter} fiber before (red) and after (blue) dispersion management using the array of metamaterial sheets. (The red curve is normalized to the amplitude of the output pulses.)}
\label{fig:pulse}
\end{figure}

The resulting pulse shape for the optimal case of Fig.~\ref{fig:disp}(c) is shown in Fig.~\ref{fig:pulse}(a). The dispersion-compensated pulse (blue line) is compared with the initial Gaussian (black circles) and the broadened/dispersed pulse (red line). The inset of Fig.~\ref{fig:pulse}(a) shows the small deviation from the initial pulse shape caused by higher-order dispersion, although it is clear from the picture that the Gaussian shape is well preserved in spite of relatively high third-order dispersion.

Finally, Fig.~\ref{fig:pulse}(b) shows a random pulse train launched into the fiber. The red curve plots the pulse train after travelling \unit{25}{\kilo\meter} inside the fiber and the blue curve plots the pulse train after passing through 130 EIT sheets. The center-to-center separation of pulses in this case is \unit{75}{\pico\second}. There is significant inter-symbol interference (overlapping of pulses) in the dispersed signal, but because of the excellent dispersion compensation by the metamaterial sheets, inter-symbol interference is completely removed in the compensated final pulse train. While the form and bandwidth of the pulses are almost perfectly restored to their initial values, the amplitude is damped to 63.9 percent of the initial pulse amplitude, still a very good result unachievable with other GVD compensation methods.
In conclusion, we have demonstrated a proof-of-principle of a dispersion compensation system using phase-engineered metamaterials providing a highly customizable dispersion band. The system can be fabricated in a compact volume using nano-fabrication methods and can be easily integrated into the communication line. The phase-engineered metamaterial can provide strong group-velocity dispersion without being adversely affected by large transmission loss. Higher-order dispersion introduced by the system is in trade-off with dissipation, and one can be exchanged for another depending on the specific line requirements.

Work at Ames Laboratory was partially supported by the U.S. Department of Energy, Office of Basic Energy Science, Division of Materials Science and Engineering (Ames Laboratory is operated for the U.S. Department of Energy by Iowa State University under contract No. DE-AC02-07CH11358), by the U.S. office of Naval Research, award No. N00014-10-1-0925 (Simulations). Work at FORTH (theory) was supported by the European Research Council under the ERC advanced grant No. 320081 (PHOTOMETA).

\bibliography{References}
\end{document}